# Multi-agent System Design for Dummies

# [A Book Draft]

By

# Li Siyao

23 Jan 2016

# ABSTRACT

Agent technology, a new paradigm in software engineering, has received attention from research and industry since 1990s. However, it is still not used widely to date because it requires expertise on both programming and agent technology; gaps among requirements, agent design, and agent deployment also pose more difficulties. Goal Net methodology attempts to solve these issues with a goal-oriented approach that resembles human behaviours, and an agent designer that supports agent development using this philosophy. However, there are limitations on existing Goal Net Designer, the design and modelling component of the agent designer. Those limitations, including limited access, difficult deployment, inflexibility in user operations, design workflows against typical Goal Net methodology workflows, and lack of data protection, have inhibited widespread adoption of Goal Net methodology.

Motivated by this, this book focuses on improvements on Goal Net Designer. In this project, Goal Net Designer is completely re-implemented using new technology with optimised software architecture and design. It allows access from all major desktop operating systems, as well as in web environment via all modern browsers. Enhancements such as refined workflows, model validation tool, access control, team collaboration tool, and link to compiler make Goal Net Designer a fully functional and powerful Integrated Development Environment. User friendliness and usability are greatly enhanced by simplifying user's actions to accomplish their tasks. User behaviour logging and quantitative feedback channel are also included to allow Goal Net Designer to continuously evolve with the power of big data analytics in future. To evaluate the new Goal Net Designer, a teachable agent has been developed with the help of Goal Net Designer and the development process is illustrated in a case study.

The new Goal Net Designer has made significant impacts. High accessibility, team collaboration, and simplified actions greatly increase efficiency of agent development. Meanwhile, powerful assistance provided in this tool ensures models are robust and well defined, and even less technical-inclined people can participate in agent design. With such enhancements, the new Goal Net Designer facilitates agent design and implementation not only in research laboratories, but also in industrial environment.

**TABLE OF CONTENTS**

# LIST OF FIGURES

# LIST OF TABLES

# 1 INTRODUCTION

## 1.1 Motivation

Agent emerged as a concept in Artificial Intelligence in late 1970s and began to be widely used from 1980s. Since early 1990s, agent technology has received considerable attention in both academic research and industry [1]. Agent technology represents a new software engineering paradigm [2], and it provides new prospect of analysing, designing, and building software systems [3]. It has also been recognised as a promising approach to develop software with characteristics that can satisfy increasing expectation of users. In early 2000s, a move from object-oriented (OO) modelling paradigm to agent-oriented (AO) modelling paradigm was observed by some researchers [4] [5]. The fundamental difference between the two paradigms is the pro-activeness of agents [5], although there are similarities between them.

Despite the early trend and research efforts in academics, widespread deployment of agent systems in industry is not seen so far [6]. Most research work only focuses on agent theories like agent mental state model, which provides theoretical foundation of agent systems. However, there are many challenges in designing and implementing real agent systems, such as gaps between requirements and agent design, gaps between design and implementation, and gaps between traditional system and agent systems [7]. Therefore, complete agent development methodologies and accompanying development tools are highly needed to bridge the gaps.

Recognising this need, researchers have proposed several methodologies for agent development, and some agent development tools have been made available. Most methodologies, such as Gaia [8], Tropos [9], and MaSE/O-MaSE [10] [11], are either extended from OO paradigm or knowledge engineering (KE) [12] [13]. Therefore, they require expert knowledge on AO paradigm, OO paradigm, or KE. In terms of popular and well-established agent development tools, JADE [14] and JACK [15], both programmed in Java, require expert knowledge both in Java and AO paradigm; PDT does not automatically correlate design data with implementation [16].

Generally, these methodologies and tools assume expert knowledge, which inhibits widespread adoption in the industry.

Among existing methodologies, Goal Net methodology requires less expert knowledge by utilising a goal-oriented approach that resembles human behaviour [17]. Therefore, it reduces entry barrier of designers and has the potential to include less technical people in agent design. Similar to most methodologies, an accompanying agent design tool was already proposed to allow agent designers to develop agents with Goal Net [16]. However, this tool has not gained wide attention due to several drawbacks that will be discussed in section 2.3.2. Therefore, to effectively enable agent development using Goal Net methodology, major improvements on agent design tool are needed. This project proposes and implements massive improvements on agent design and modelling tool, Goal Net Designer, to simplify agent development and connect agent design with implementation.

## 1.2  Problem Statement

The overall aim of this project is to re-develop and improve the agent design tool, Goal Net Designer, as an Integrated Development Environment (IDE) so that it can be open to developers in industry and researchers around the world. Detailed objectives of this project are:

- To redesign and redevelop Goal Net Designer which realises agent design using Goal Net methodology. This IDE should be intuitive to use, cross-platform, and easily accessible by users.
- To develop a Goal Net model compilation module in Goal Net Designer. This model compilation module should be able to validate a Goal Net model, and link IDE to appropriate runtime executable (compiler).
- To design and develop team collaboration, sharing, and access control modules in the IDE so that it enables teamwork and communication needs in industrial environment.

- To record user behaviours and feedback in the IDE to enable big data analytics in future.

## 1.3 Contributions

Main contributions of this project include:

- A fully-functional IDE for Goal Net methodology. This new IDE allows desktop access from all major desktop operating systems and web access from all major browsers. With new architecture and improved design, it has greatly enhanced flexibility and user experience.
- A Goal Net compilation module. This rule-based module validates Goal Net model and classifies feedback into errors and warnings. It allows user to easily locate problems with current design. It also allows execution of current Goal Net by triggering external compilers.
- Team collaboration, sharing, and access control modules. These modules altogether allow Goal Net modelling in a collaborative team and enable reuse of basic components in Goal Nets.
- User behaviour tracking and feedback collection module. User feedback and behaviours are stored in a centralised database with appropriate granularity for big data analytics in future.
- A case study to demonstrate the development of agent application using Goal Net methodology and proposed IDE with greater efficiency and productivity.

## 1.4 Report Organisation

Following this introduction, the remainder of this report is organised as follows. Section 2 describes related theories on agents, Goal Net methodology, and existing Multi-Agent Development Environment. Section 3 starts with requirements of the project based on user feedback, followed by new software architecture and database design. It also explains the new features and major enhancements in detail. Section 4 presents a demonstration on how this IDE fits into agent development process with great usability. Section 5 presents a case study where the IDE is used in development of a teachable agent. Section 6 concludes this report and discusses the future work.

## 2 THEORETICAL BACKGROUND AND RELATED WORK

This section consists of three sub-sections. Section 2.1 covers background on software agent and agent-oriented software engineering. Based on understanding of software agent, section 2.2 illustrates Goal Net, the theoretical foundation of this project. In section 2.3, the existing development tool for Goal Net and its drawbacks are discussed.

### 2.1 Software Agent

#### 2.1.1 Agent and Its Goal Orientation

Although used frequently in many research areas, there is no universal definition of software agent in the context of computer science. Many definitions of software agents have been proposed by various researchers, but none of them has received widespread acceptance [7]. Instead, researchers usually regard agent as a software-based computer system that has the following properties [18]:

- Reactivity: agents are able to perceive external environment and timely react to environmental changes [18].
- Goal orientation (pro-activeness): in addition to reactivity, agents are able to exhibit goal-directed behaviour by taking initiative to plan and execute.
- Autonomy: Agents can have some control on its actions and internal states, and they operate without direct supervision of and help from human being. It is tied closely to pro-activeness.
- Social ability: agents are able to interact with each other or humans. It enables agent collaboration especially in multi-agent system, and it is usually achieved through predefined agent communication language [1].

Among the four commonly accepted characteristics, goal orientation is regarded as the most important one as it is the main difference between agent and object. It has also been recognised as a paradigm in modelling agent-based system [7]. Moreover, many researchers believe that goal-oriented agents are able to exhibit goal autonomy, with which agents are able to select reachable goals based on their own mental states and perception of external environment. Goal Net methodology captures goal-oriented behaviours of agent, and therefore, agents designed by Goal Net Designer are able to exhibit goal autonomy.

### 2.1.2 Agent-oriented Software Engineering

Currently, most of the existing agent system are still developed using OO concept. However, given the goal-oriented nature of agents, a shift from OO paradigm to AO paradigm will be beneficial to agent design. The concept of Agent-oriented Software Engineering (AOSE) emerged from this shift. From AOSE point of view, a software is considered a multi-agent system, in which autonomous and de-centralised agents interact and pursue their own goals. Goal Net Designer, as an AOSE development tool, is used to design individual agent that can be used alone or in multi-agent systems.

## 2.2 Agent Development Using Goal Net

### 2.2.1 Goal Net Model

Goal Net is a goal-oriented agent modelling method proposed by Zhiqi Shen [17]. As a composite goal model, it is used as a mental model for goal-oriented agent. A Goal Net model has three basic components:

- State. States are used to represent intermediate and final goal states that an agent should achieve.
- Transition. Transitions determine the relationship between states. Each transition has at least an input state and an output state.
- Arc. Arcs are directed associations that connect state to transition, or transition to state.

A simple example of Goal Net model consisting of all components mentioned above are given in Figure 2.1. It describes a transition k connecting state i with state j.

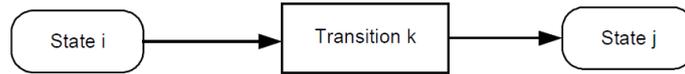

Figure 2.1    A Goal Net with basic elements [7]

A state contains attributes such as ID, name, description, type, achievement value, and cost. There are two types of state, namely atomic (simple) state and composite state. Atomic state cannot be split, but a composite state can be split into states connected via transitions. Behaviours of agent in a state can be defined by a list of functions that are associated with this state.

Similarly, a transition contains attributes including ID, name, description, and type. There are three types of transitions. Direct transition performs transition from input state to output state via a fixed task execution. Conditional transition performs a rule-based reasoning to select target output state. Probabilistic transition is used in uncertain environment where probabilistic inference determines target state. Actions to be performed in a transition are defined by a list of tasks. In a task, one or more functions can be executed.

An arc contains attributes such as ID, name, description, input entity ID, and output entity ID. Input entity ID and output entity ID are either ID of a state or ID of a transaction, depending on direction of the arc.

Finally, as the highest level abstraction, a Goal Net has a number of properties. Root state is the root of Goal Net, and it must be a composite state. Start state represents the initial state that the agent is in, and end state is the goal state that the agent strives to reach.

Using above-mentioned hierarchical state structure to represent different levels of goals and transactions to model goal orientation of agent, Goal Net model is able to handle the complexity in agent design. Figure 2.2 shows a complete Goal Net with different level of goals and different transitions.

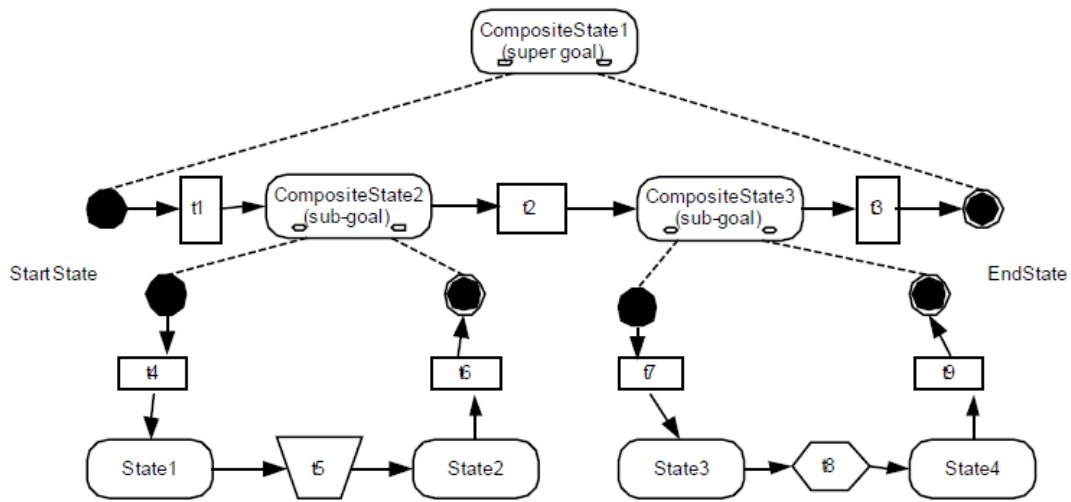

Figure 2.2    A complete Goal Net with different types of states and transitions [7]

### 2.2.2  Goal Net Methodology

Built on Goal Net model, Goal Net methodology is a complete guide and process to design and implement agents using Goal Net model. As an AOSE methodology, it covers the entire development lifecycle of an agent system. It has some similarities to traditional Software Development Life Cycle (SDLC), but it has a stronger focus on agent design and modelling. There are four phases in Goal Net methodology:

- Requirement Analysis: In this phase, system requirements are gathered to design a top level Goal Net. This Goal Net serves as an overall description of the system and will be refined in subsequent phases.
- Agent Architecture Design: Goal Net from Requirement Analysis phase is split into sub Goal Nets. Multiple agents will probably emerge in this phase, and each of them will incorporate a sub Goal Net.
- Detailed Design: Goal Net for each agent is refined, and agent behaviour is determined by linking states to functions and transitions to tasks.
- Implementation: Detailed designs of agents are used to generate agent skeleton, and programming work is involved to implement functions that determine agent behaviours. This phase involves little knowledge prerequisite on AO paradigm and AOSE.

Goal Net Designer in this project supports the first three phases of this methodology, and therefore exhibits strong AOSE characteristics. It also connects design with implementation by capturing data that will be helpful in the implementation phase.

## 2.3 Multi-Agent Development Environment (MADE) for Goal Net

### 2.3.1 MADE Architecture

To support agent development using Goal Net methodology, MADE employing Goal Net methodology has been developed [16]. It includes an existing Goal Net Designer, an agent creator, and a knowledge loader.

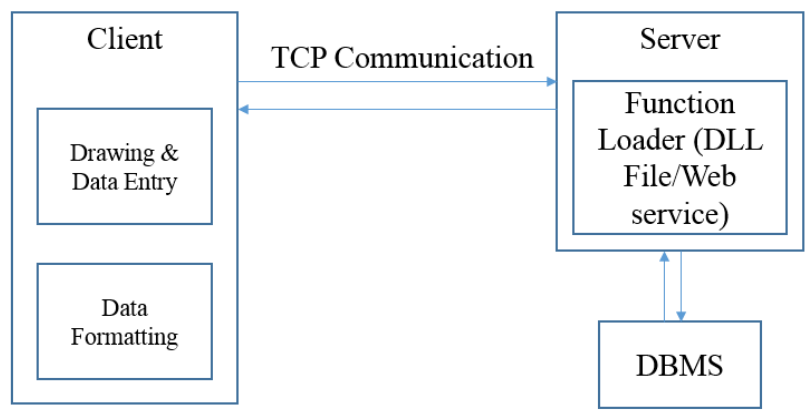

Figure 2.3    Architecture of previous Goal Net Designer

Among the three components of MADE, Goal Net Designer is the most crucial one because of not only its AO paradigm, but also its important role in all phases of agent development. The existing Goal Net Designer employs a client-server architecture as shown in Figure 2.3. All Goal Nets are stored in database located at server side, which also maintains a list of available functions that are loaded from DLL files or web services. Client side connects to server to load Goal Net and allows Goal Net drawing via a GUI. The other two components essentially act as an interpreter, where agent creator creates agents and knowledge loader injects Goal Net design data into each agent.

This process establishes a complete workflow in building agents. However, up to now, it only receives limited use in few research laboratories due to some significant drawbacks in both system design and system features. Those drawbacks are discussed in detail in next section.

### 2.3.2 Major Drawbacks of Existing MADE

Major drawbacks of existing MADE mainly lie in Goal Net Designer, and those drawbacks can be classified into two categories by their direct consequences, namely impeding widespread adoption of this MADE and reducing development efficiency as well as user experience.

The following two factors impede widespread adoption of MADE:

- The existing Goal Net Designer is not easily deployable. To begin modelling with Goal Net Designer, one must be able to access a deployed instance of it. However, the client-server architecture poses many extra steps and problems to deployment. A database instance and a server instance must be deployed prior to installation of client programme. Many manual setups in configuration files are required to establish the connection between client and server. Also, each update of Goal Net Designer will require a redeployment, which again adds on to the burden of end user. These problems on deployment and installation definitely affect an agent designer's decision. Moreover, it requires some technical knowledge to set up database and a client-server connection; therefore, less technical-inclined people will face difficulties when trying to start with the existing Goal Net Designer.
- Past implementation only supports desktop access on Windows platform. When the tool was developed in 2007, this was probably not a major issue. However, given the development of web technology and the increasing popularity of Unix-based operating systems, a tool lacking cross-platform access and web access will have difficulties in attracting users.

Additionally, users also experience many inconveniences while using the existing Goal Net Designer:

- The existing Goal Net Designer does not offer great usability. Firstly, the existing Goal Net Designer does not support change of state type. In other words, no conversion between atomic state and composite state is allowed. This poses great inconvenience in agent development because one of the major steps in Goal Net methodology is to split goals into sub-goals and assign them to different agents. Secondly, moving and deleting a drawn object, although used frequently, is not intuitive. Users have to select "Move Mode" or "Delete Mode" prior to selecting an object of interest. Moreover, lack of support to move or delete a group of objects causes big trouble when dealing with large Goal Nets. Such lack in functionality and action sequences severely affects development experience and efficiency.
- There is little flexibility while dealing with functions used in Goal Net. The existing Goal Net Designer requires functions to be stored in the server database prior to using it in a Goal Net. Moreover, it only supports importing function from DLL files and web services but not manual entry. This mode of work causes two problems. Firstly, function interfaces are forced to be ready before design, but according to Goal Net methodology, such action should only commence in implementation phase due to frequent changes in design data. Secondly, communication overhead between server manager and client user is dramatically increased, and design work may have to pause before required function is added at server side.
- Little assistance is provided for Goal Net modelling. Although some feedback is given during drawing, there is no indication of source of the error nor advice to fix the problem. Also, user has no way to ask the programme to check if the model is correct, after user thinks they have fixed the error. This introduces troubles to people who are not familiar with Goal Net.

- Work done in existing Goal Net Designer is easily tampered. There is no access control in existing Goal Net Designer, so anyone with a client programme can connect to server and modify existing work. Moreover, functions are shared across all Goal Nets stored in server and therefore, changes in a function may affect multiple Goal Nets without any notification. Also, there is no local buffer for a Goal Net so whatever changes made to a Goal Net is final. These factors harm integrity of Goal Net design data, and may cause serious trouble in implementation of agents.

Given the drawbacks discussed above, complete overhaul to the existing Goal Net Designer is inevitable.

# 3 IMPLEMENTATION OF NEW GOAL NET DESIGNER

This section, consisting of 4 sub-sections, discusses implementation details of new Goal Net Designer. Section 3.1 specifies the scope of work based on previous discussion on drawbacks and user feedback collected. Section 3.2 presents new architecture and design of Goal Net Designer. Section 3.3 presents new features added to new Goal Net Designer and their mechanisms. Section 3.4 briefly describes testing and deployment techniques.

## 3.1 Scope and Requirements

Target of enhancement in this project is determined to be Goal Net Designer in MADE because of its importance and drawbacks discussed above. After careful examination of previous code and feedback collected from users in Joint UBC-NTU Research Centre of Excellence in Active Living for Elderly, a decision is made to completely re-implement Goal Net Designer. This decision is made because:

- Previous code written in C# as Windows Form application is obsolete and naturally inhibits cross-platform and web access;
- Efforts spent to reuse existing code are not worth the rewards, given dramatic changes foreseeable;
- Previous code is not easily maintainable and understood

Scope of re-implementation of Goal Net Designer and further improvements is determined to include:

- A fully-functional IDE which allows cross-platform access and web access, written in Java using JavaFX
- Greatly improved workflows with easy-to-use edit functions, group operations, and removal of linking function to DLL file and web service
- Team collaboration and access control
- Goal Net model checking module and integration with compiler
- User behaviour logging and feedback channel

For simplicity, the word Goal Net Designer refers to the new implementation of Goal Net Designer thereafter unless otherwise specified.

## 3.2 Architecture and Design

In this section, software level architecture and class-level architecture, together with design considerations, are discussed. This section also includes a database design to capture Goal Net design data and enable all features required.

For better presentation, this section only includes class diagrams that are necessary for illustration.

### 3.2.1 Software Architecture

As discussed in section 2.3.2, there are some drawbacks directly associated with client-server architecture. Therefore, client-server architecture is abandoned. In the new architecture, there is no server side programme, although a shared remote database is still required for data storage and data sharing purpose. The new architecture is given in Figure 3.1.

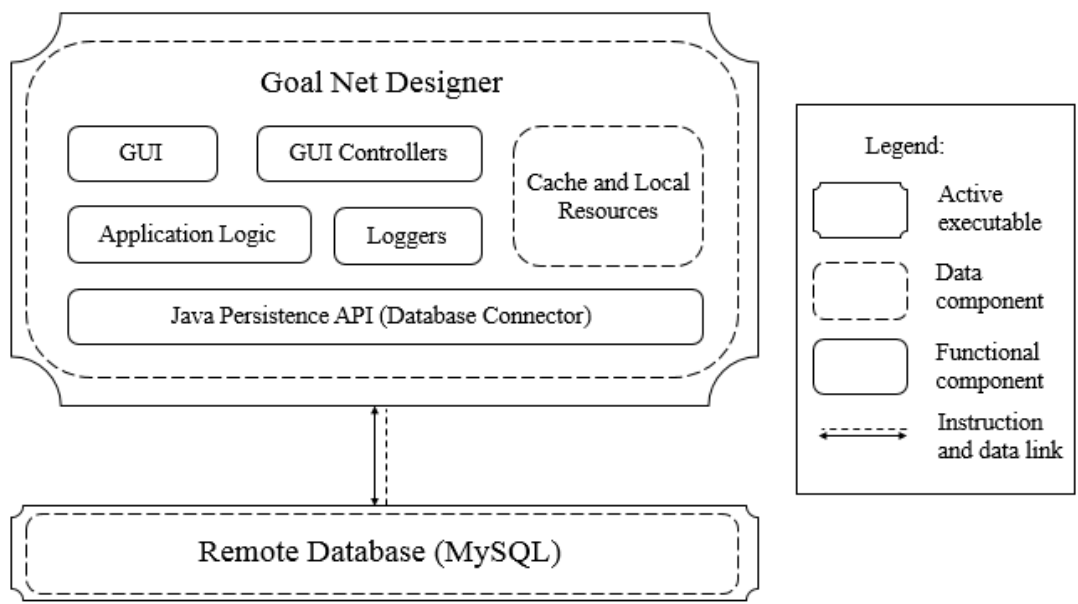

Figure 3.1   New architecture of Goal Net Designer

In the new architecture, server side programme is completely removed and client side directly communicates with database. Data integrity will be ensured by application logic and Java Persistence API (JPA).

Goal Net Designer at client side is divided into several interconnected components. GUI and GUI controllers are in charge of look and feel of user interface. They also contain necessary classes for display of each Goal Net entity. Application logic includes all middle-level managers that are in charge of control logics, data integrity check, and local cache management. Loggers enable logging to database and GUI.

Communication with database is completely handled by JPA. JPA is a Java framework which allows object-relational mapping (ORM) between Java classes and database tables. For each database table and composite primary key, a corresponding Java class is created. When communicating with database, a JPA transaction maintains status of each managed class, and thus allows management of database by simply dealing with Java classes. In this way, SQL queries are completely erased from code and thus make the code more robust. Also, built-in tool which allows constructing class based on existing database greatly simplifies code modification upon changes to database structure.

This new architecture, together with accompanying deployment strategy discussed in section 3.4, successfully resolves the drawbacks of previous client-server architecture. Now, to use Goal Net Designer, users will only need to launch client side programme. The launch can be performed in all major desktop environments, or in all major web browsers by visiting a public web URL that will be established in deployment. To start development instantly, user can connect to a shared database which will be set up when this designer is released for public access. They also have the option to have a local database if they have enough expertise.

### 3.2.2 Class-level Design: Model-View-Controller and Generics

One of the concerns when deciding if previous code should be used is that the previous code does not have a clear structure and classes are tightly coupled.

Therefore, reusability of existing code is quite low, which makes future improvements on the code base difficult and inefficient. Considering the fact that Goal Net designer will certainly require further updates and bug fixes, new implementation should have clearly defined class structure which is loosely coupled and easy to maintain. Since GUI is heavily involved in Goal Net Designer, model-view-controller (MVC) style becomes a natural choice.

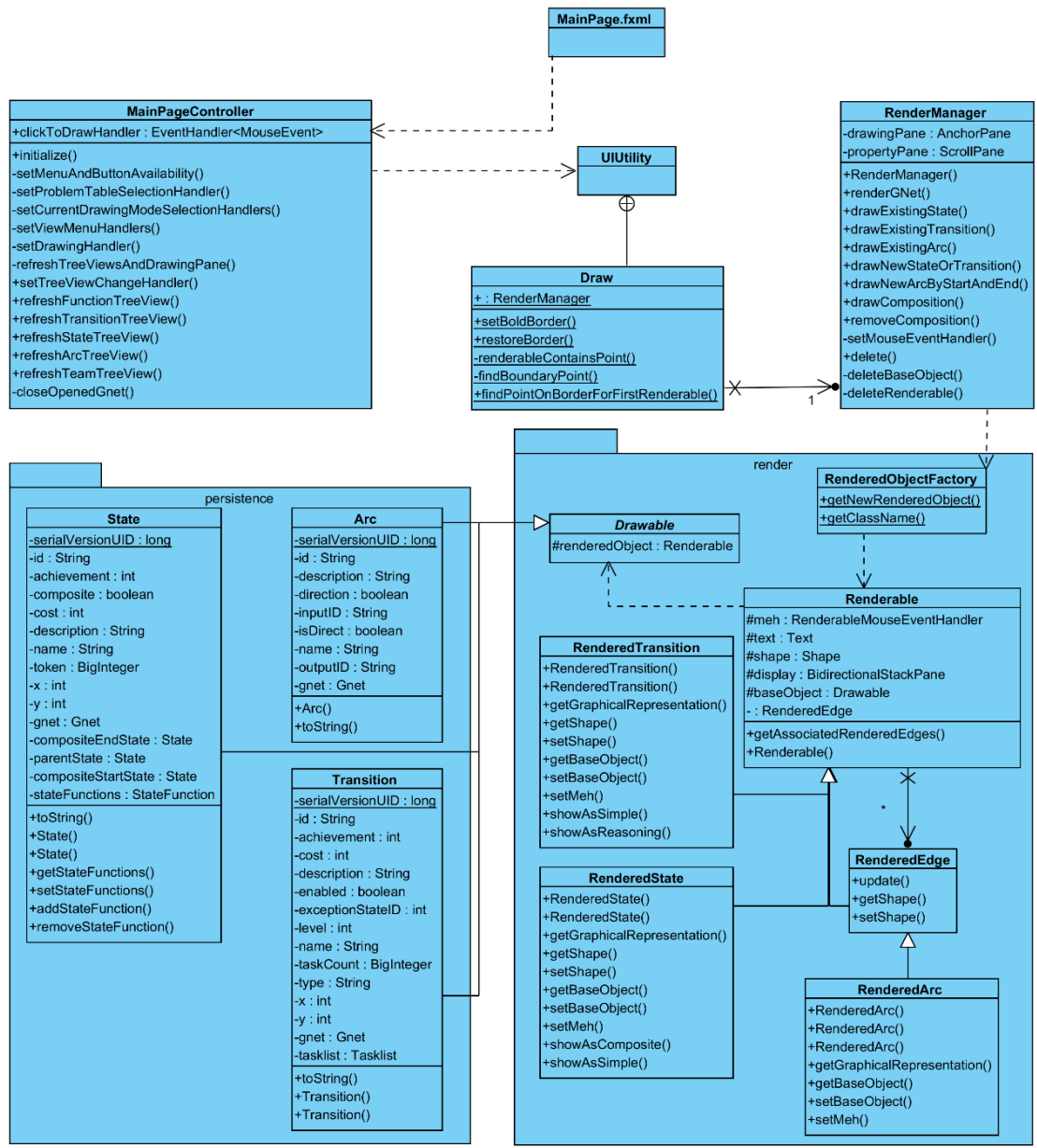

Figure 3.2    Example of MVC design in Goal Net Designer

Figure 3.2 gives an example of MVC design in Goal Net Designer. This class structure is used for displaying states, transitions, and arcs in Goal Net Designer. For simplicity, getters, setters, and all unnecessary classes are omitted.

In the example, MainPage.fxml is the JavaFX Markup file which defines GUI view. MainPageController is the direct controller of GUI. Model, according to generally accepted definition of MVC, consists of functionality and data. In Figure 3.2, RenderManager, render package, and persistence package are considered as model in MVC design. By doing so, GUI is loosely coupled with core functionality, and it makes the code easily maintainable.

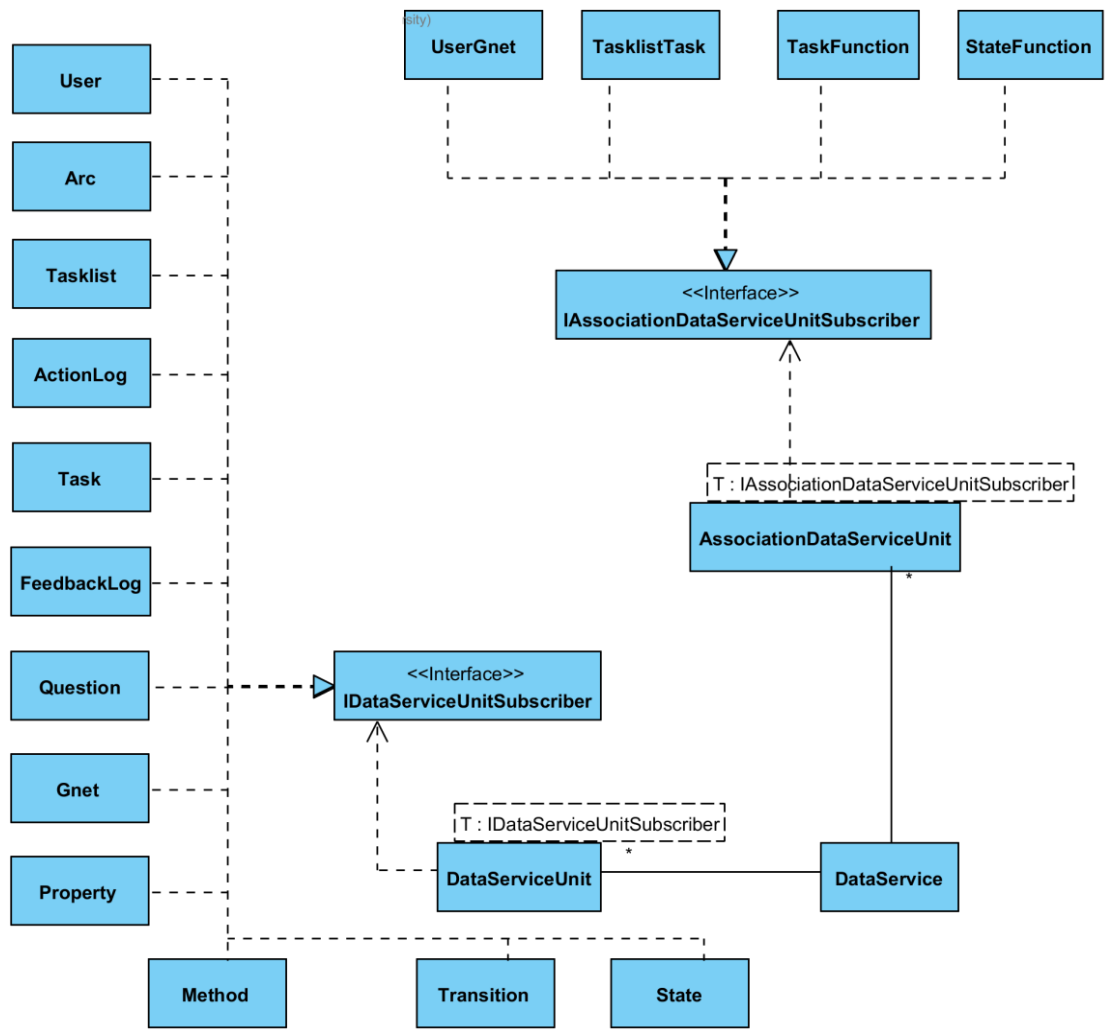

Figure 3.3    Generic classes in database access control
XVII

Another effort to reduce code size and increase maintainability is the use of generic classes when dealing with database access. Traditional way to manage an object in database is to create one Data Access Object for each entity class to manage creation, read, update, and deletion of the object. However, it results in repetition in coding as most operations are the same. To minimise code repetition, generic class is introduced.

Figure 3.3 illustrates the implementation of database access control part in Goal Net Designer. DataServiceUnit<T> is in charge of all database operations for type T, where T is one of the simple entities. AssociationDataServiceUnit<T> is the generic class for database operations related to all association classes that are mapped to association tables in database. The two generic classes are separated because of different implementations for some database operations. DataService contains one singleton instance of DataServiceUnit<T> or AssociationDataServiceUnit<T> for each entity. Database operations on an object are achieved by triggering methods in corresponding instance in DataService class.

With above-mentioned class-level designs, the new code base can be easily maintained, understood, and extended. This reduces the efforts required for future updates and the possibility of complete re-implementation similar to this project.

### 3.2.3 Database Design

As an integrated part of Goal Net Designer, database should be able to store all design data for all Goal Nets. Additionally, supportive database tables are required to realise user feedback channel, user behaviour logging, and access control. Figure 3.4 shows the database structure for Goal Net Designer, and those database tables can be classified as follows:

- Tables for Goal Net data: gnet, state, transition, arc, method, tasklist, and association tables between these tables
- Tables for feedback channel: question, feedback_log, and user
- Tables for user behaviour logging: action_log, and user

- Tables for access control: user and user_gnet

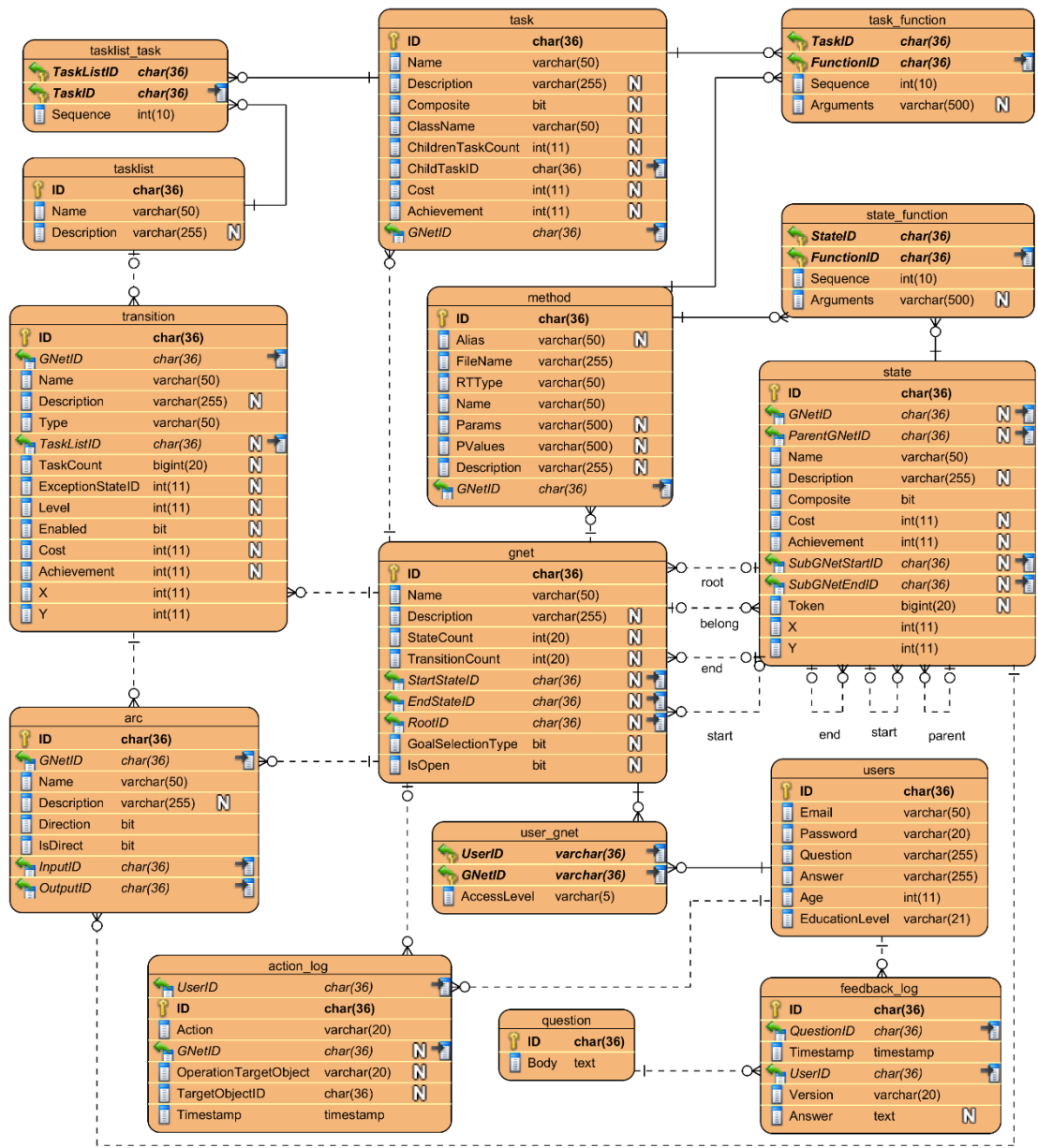

Figure 3.4    Database design of Goal Net Designer

All primary key fields, except the one in user table, use universally unique identifier (UUID) to reduce complexity in database merging. The exception is made because user needs an ID to login, but UUID is hard to remember. Also, in the case where another unique identifier is given to user, there might still be conflicts. Therefore, a

XIX

compromise has been made that in the case of user ID conflicts during database merging, some quick manual edits using simple SQL queries are still required.

In summary, the new software architecture, class-level design and database have helped with getting rid of problems in previous implementation. All features for Goal Net modelling are developed based on designs above. The next chapter discusses design and implementation of additional features in Goal Net Designer.

### 3.3 Additional Features

This section discusses working mechanisms of new features in detail. Demonstration and screenshots for these new features are presented in section 4.

### 3.3.1 Access Control and Team Collaboration

To control user's access to a Goal Net, three levels of access are designed with corresponding rights as listed in Table 3.1 below.

| Access Level | Access Right |
| --- | --- |
| Admin | Write Access + control user access rights of Goal Net |
| Write | Read Access + edit Goal Net and save to database; share task and function |
| Read | view Goal Net, edit without saving to database, and export as PNG file |

Table 3.1    Access levels and corresponding access rights

Creator of a Goal Net will automatically have Admin Access to it. A Goal Net can only be opened by users who at least have Read Access to it.

Access control ensures proper management of access to Goal Net, but on the other hand, it may result in repetitive work because there might be some scenarios where existing Goal Net designs, especially task and functions, can be reused in another Goal Net. To ease this problem and allow more efficient team collaboration, two clone features are incorporated to assist sharing of design data:

- Clone function: this feature allows user to clone a function from Goal Net A to Goal Net B. User must have at least Write Access to both Goal Nets.
- Clone task: this feature allows user to clone a task from Goal Net A to Goal Net B. If this task contains functions, all functions in this task and their associations with the task will be cloned, too. User must have at least Write Access to both Goal Nets.

New entries will be created in database for both clone operations. Therefore, editing the new entity will not affect original entity.

With access control scheme and clone functions discussed above, Goal Net Designer can be used in a team environment and multiple users can collaborate on a Goal Net. Design reuse is also enabled while maintaining data integrity. This new feature will increase efficiency of agent development and greatly reduce communication overhead.

**3.3.2  Goal Net Model Validation**

To design a robust agent using Goal Net methodology, a robust agent design in Goal Net is crucial. Goal Net model validation tool in Goal Net Designer acts an assistant to help user identify problems in Goal Net. In the context of a compiler, the model validation tool is a syntax analyser for Goal Net based on structured data in database.

The model validation tool is implemented as a rule-based validator. Problems on Goal Net are classified into errors and warnings by seriousness. Errors are problems which must be fixed before this Goal Net can be used in the next phase. Warnings are problems which may cause agent to behave in an unintended way. The following rules are used to identify errors:

1. A Goal Net has no root state, start state, or end state
2. A Goal Net's root state is not a composite state
3. A composite state has no start state or end state
4. A non-root state is not connected to any transition
5. A transition is not connected to any state
6. A transition has only outgoing arc, or only incoming arc

The following rules are used to identify warnings:

1. A state has no associated function
2. A transition has no associated task
3. A task has no associated function
4. A state has only outgoing arcs but it is not start state
5. A state has only incoming arcs but it is not end state

Figure 3.5 shows the class diagram related to this validation tool. To make adding and removing rules easier, actual implementation classifies rules by type of entity. For example, rules related to state are stored in StateValidator, and rules related to transition are stored in TransitionValidator. ValidationManager maintains a list of validators and will trigger validate function in each of them when validation starts.

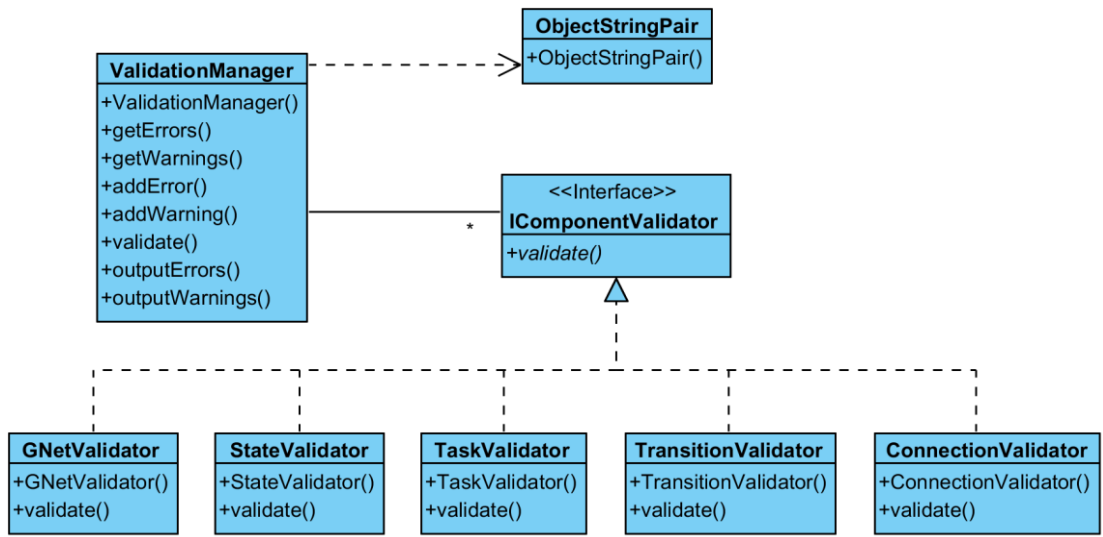

Figure 3.5   Class structure of Goal Net validation tool

Additionally, a special helper class ObjectStringPair is used to associate problem description with source of the problem. With this information, user will be directed to source of the problem automatically when they select to investigate a problem, and they can quickly fix the problem. It further helps less technical-inclined users in agent development.

### 3.3.3   Link to External Compiler

To bridge the gap between design phase and implementation phase, the path of compiler can be specified by user to run current Goal Net in Goal Net Designer. When user selects to run current Goal Net, Goal Net validation tool will firstly be triggered. If there is no error, Goal Net Designer will start the specified compiler and pass UUID of current Goal Net to it. However, if there are any errors, user will be asked to resolve the errors before launching.

This link successfully resolves the trouble of manual start of agent creator and knowledge loader, and it is the final building block that makes Goal Net Designer a real IDE.

### 3.3.4   User Behaviour Tracking and Analytics

In Goal Net Designer, user behaviours are tracked by a logger that is injected into every user operation once a Goal Net is opened. Table 3.2 shows all objects of interest, actions tracked for each type of object, and data collected in each entry of user behaviour log.

Although trivial in implementation, the impact of this logger is significant. Currently, big data analytics is a hot topic in computer science. Tracking of user behaviours introduces the possibility of using big data analytics to analyse user's usage pattern in future. Based on usage pattern, it is possible to identify some common difficulties faced by user, and therefore, allow enhancements to be made continuously on Goal Net Designer. Moreover, with data on user's age and education level that are collected at registration time, data analytics may be powerful enough to reveal different user habits for different groups. It may contribute to personalised optimisation on Goal Net Designer, and may generate some research insight, too.

| Object Type | Actions Type Tracked | Data Collected |
|---|---|---|
| Goal Net | Open, Close, Edit | Object type, Object ID, User ID, Action type, Timestamp |
| State and Transition | Create, Edit, Move, Delete | |
| Arc, Function, Task, and Association between: function and state, transition and task, task and function | Create, Edit, Delete | |

Table 3.2     Object type, action type, and data collected by built-in user behaviour tracker in Goal Net Designer

### 3.3.5    Official Feedback Channel

On top of user behaviour tracking, Goal Net Designer also incorporates an official feedback channel which contains a questionnaire. It is used to investigate user's satisfaction with current version of Goal Net Designer. To enable quantitative evaluation on each version, questions are all answered with a 5-point scale slider. The official feedback channel does not constitute a field for custom advice because they are meant to be evaluated on case-by-case basis.

This feedback channel is specially designed to ensure flexibility and configurability. It is foreseeable that after a version of Goal Net Designer is deployed, questions in the questionnaire may be modified. Sometimes, different questions may be used for the same version to for more accurate evaluation. Therefore, questions in feedback form must be configurable without a release or deployment. To achieve this, all questions are stored in database and dynamically loaded in Goal Net Designer.

Together with user behaviour tracking, this feedback channel provides Goal Net Designer with a comprehensive toolbox to gather feedback and points of improvements, which will certainly guide continuous enhancements in future.

### 3.4    Testing and Deployment Strategy

Unit test of Goal Net Designer were conducted immediately after implementation. After unit test, new Goal Net Designer was sent to users in Joint NTU-UBC Research Centre of Excellence in Active Living for the Elderly. This prototype was well received, and enhanced usability on moving an object and group operations were highly spoke of.

| **Deployment Mode** | **Environment** |
| --- | --- |
|  |  |

| JAR file | Desktop OS with JRE |
| --- | --- |
| Java Network Launch Protocol | Web (hyperlink) |
| Web Applet (signed) | Embedded in webpage |
| Native executable (wrapped according to targeted OS) | Desktop OS |

Table 3.3    Deployment modes and target environments of Goal Net Designer

Deployment of Goal Net Designer will be done using Ant. Goal Net Designer supports four modes of deployment as shown in Table 3.3. The rich selection will certainly be a value-added point as it allows Goal Net Designer to be accessible in most environments. Especially, web access, together with remote database on cloud, allows agent designers to work anywhere as long as they are connected to the Internet via PC. Also, as all versions are compiled from the same source, users will receive universal experience in all environments. Without platform restrictions, Goal Net Designer is now able to reach a larger community of agent designers.

## 4 DEMONSTRATION OF GOAL NET DESIGNER

This section demonstrates a typical workflow where user designs a single agent in Goal Net Designer. Section 4.1 demonstrates creating a Goal Net, drawing a Goal Net consisting of states, transitions, and arcs, and linking functions, tasks, and states. Section 4.2 demonstrates validating Goal Net model and fixing errors and warnings in Goal Net. Section 4.3 demonstrated sharing a Goal Net with teammate, and cloning function and task to another Goal Net.

### 4.1 Goal Net Modelling

To start Goal Net modelling, a Goal Net must be created as shown in Figure 4.1. Here, a plan called SDLC is created using account lisiyao.

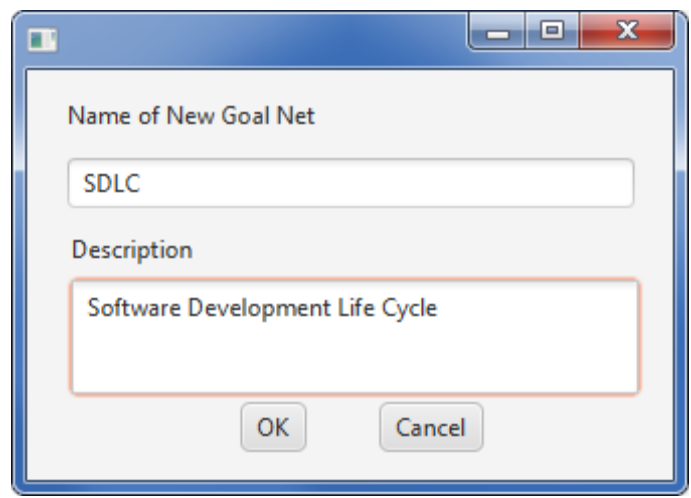

Figure 4.1    Creation of Goal Net

After creation of this Goal Net, user is directed to the main page where design can be performed. Appearance of the main page is given in Figure 4.2. It contains a menu bar on the top, followed by a toolbar to select action for drawing a Goal Net. Drawing pane is the place where Goal Net model will be constructed. Lower left corner is the standard output window of this IDE. On the upper right corner, all arcs, functions, states, tasks, and transitions are listed for easy selection. The bottom right corner contains collaboration information and allows user to edit properties of currently selected object. Currently, this Goal Net contains no drawing so it looks a bit empty.

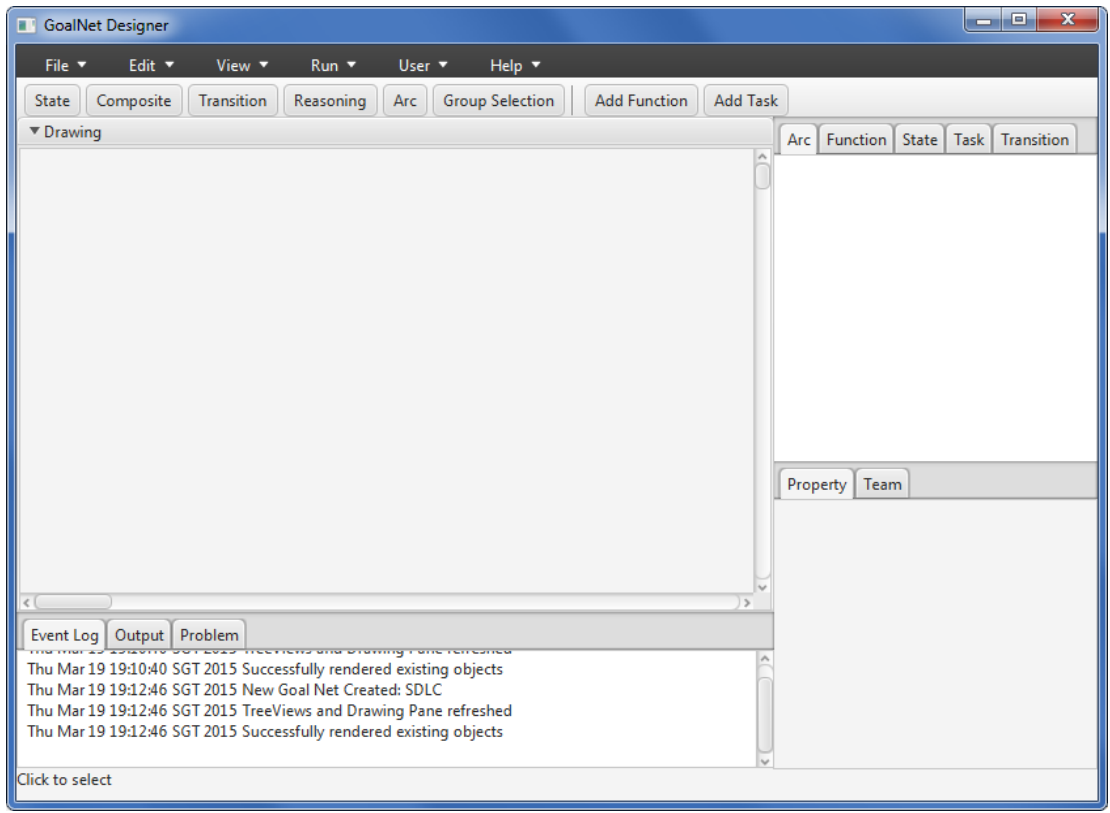

Figure 4.2    Main page of Goal Net Designer

In order to start modelling SDLC, a composite state named SDLC is drawn by selecting "composite" from the toolbar and clicking on the place where the user would like to place this composite state. The main page after these actions is shown in Figure 4.3.

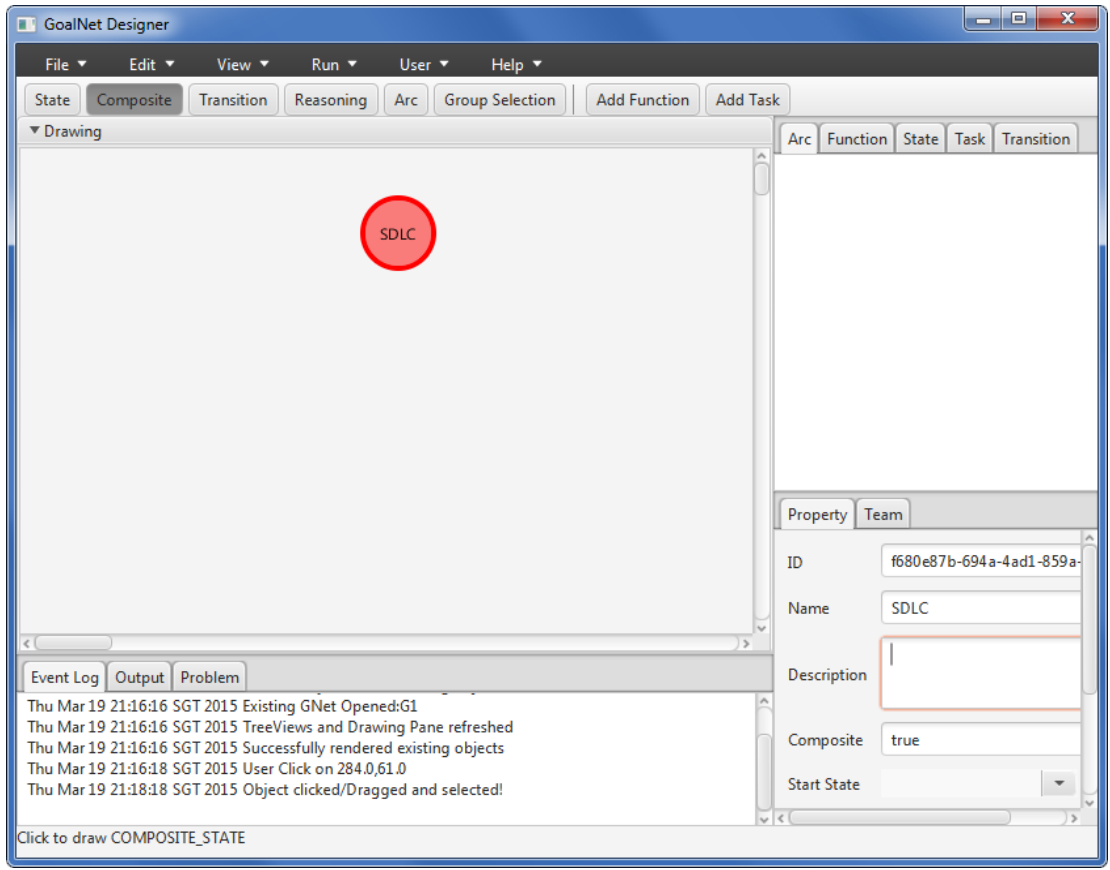

Figure 4.3    Drawing of composite state SDLC

Based on this, more states and transitions are added to model the entire waterfall SDLC. Figure 4.4 shows the completed SDLC modelling in this example. Green circles represent atomic states; arrows represent arcs; rectangles represent transitions. The dotted lines with arrow show the start and end of a composition state. In this example, the composite state SDLC starts with state "Start", and ends with state "End". For demonstration purpose, this SDLC is much simplified.

Furthermore, functions and tasks are added by "Add Function" and "Add Task" on toolbar. After that, tasks are associated with transitions and functions are associated with states. Such associations can be easily done in the property pane. In Figure 4.4, transition "Design Software" is selected, and by clicking "Manage Tasks" in property pane, a window as shown in Figure 4.5(a) pops up to manage tasks associated with this transition. Similarly, Figure 4.5(b) shows a window to manage functions associated with a state. This Goal Net is saved to database after above-mentioned edits.

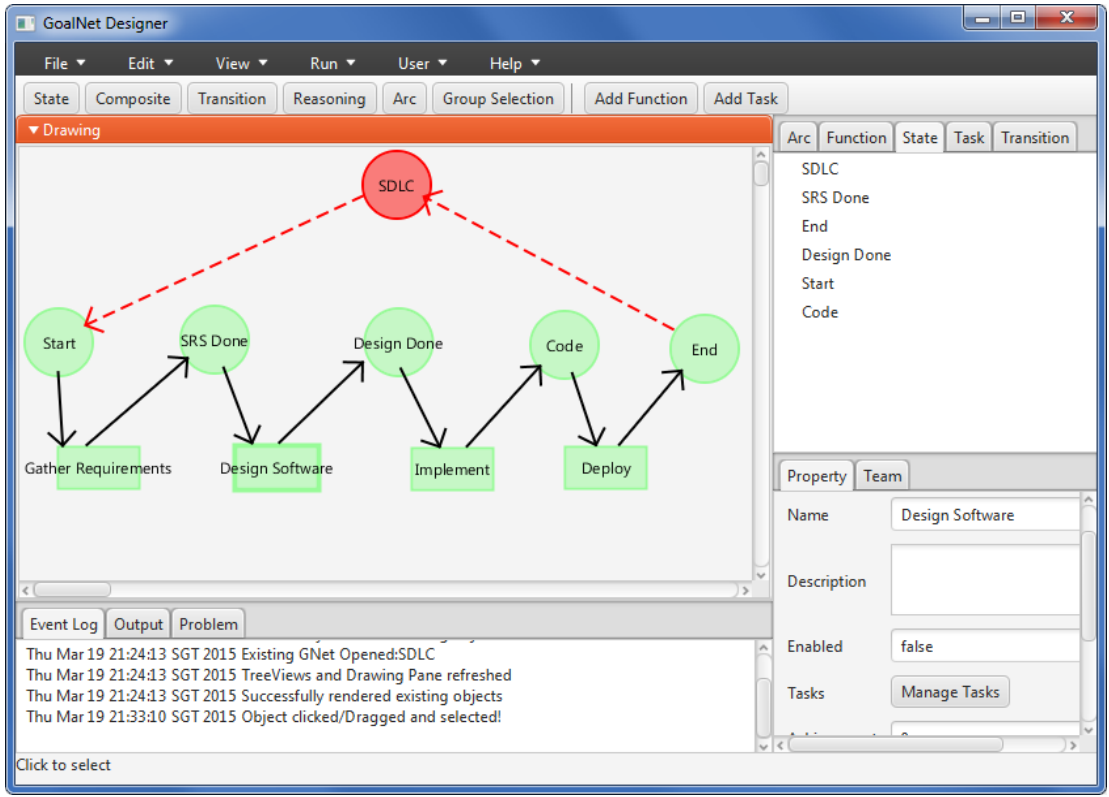

Figure 4.4    Completed Goal Net design example: SDLC

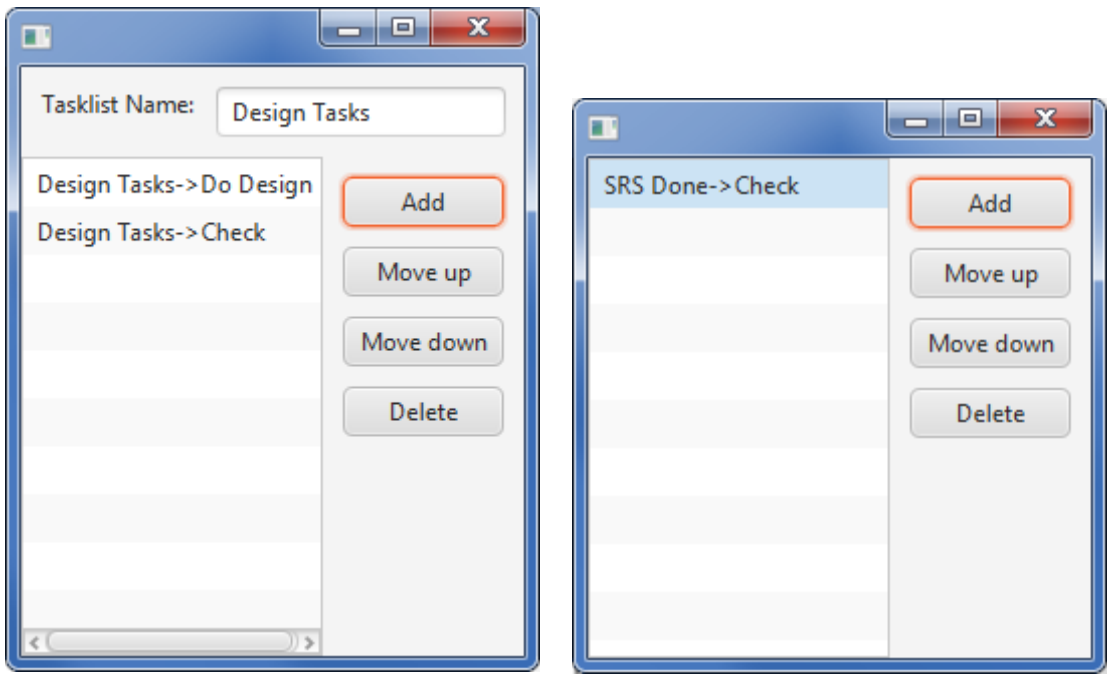

xxx

(a) (b)

Figure 4.5    Manage tasks of transition (a) and manage functions of state (b)

## 4.2 Validation and Run of Model

After drawing SDLC model in Goal Net Designer, the designer wants to check if the model is good enough for implementation. Therefore, Goal Net model validation tool is triggered by Run→Validate (F10). In this validation, warnings are ignored.

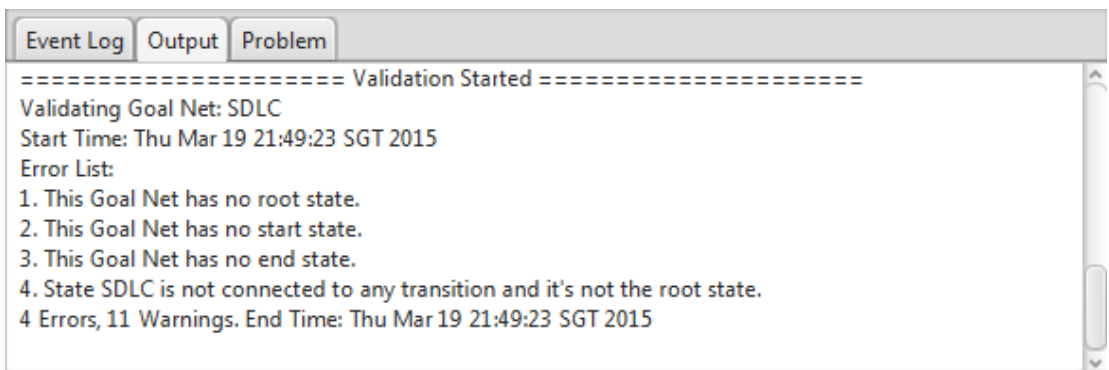

Figure 4.6    Outputs of Goal Net model validation tool

Output from validation tool is given in Figure 4.6. Four errors are given by the model validation tools. To get a more structured view, user switches to "problem" tab, where the errors are given in a list, as shown in Figure 4.7.

Figure 4.7    Problem tab after running model validation tool

To fix these errors, users should set start, end, and root state for Goal Net. Normally, it is done in Edit→Goal Net Properties (Ctrl+Shift+G). However, if user does not know how to do it, this problem tab can help by automatically directing user to the place where the error can be fixed. Here, when user clicks on the first entry, the window shown in Figure 4.8 pops up for user to fix this error.

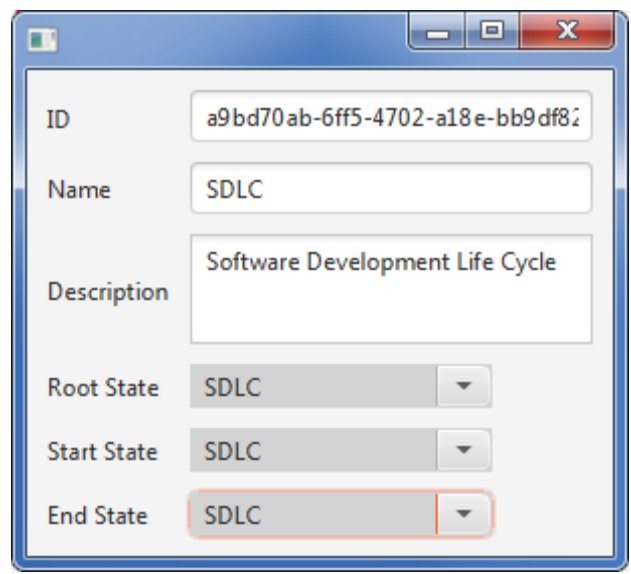

Figure 4.8    Page to edit Goal Net properties

After fixing this error, the model is good enough for running. When triggering Run→Run (F5), Goal Net Designer gives an error that external compiler is not specified. It then asks user to set the path to the executable file, as shown in Figure 4.9. After such configuration, by pressing F5 key, the model can be run by the specified executable file.

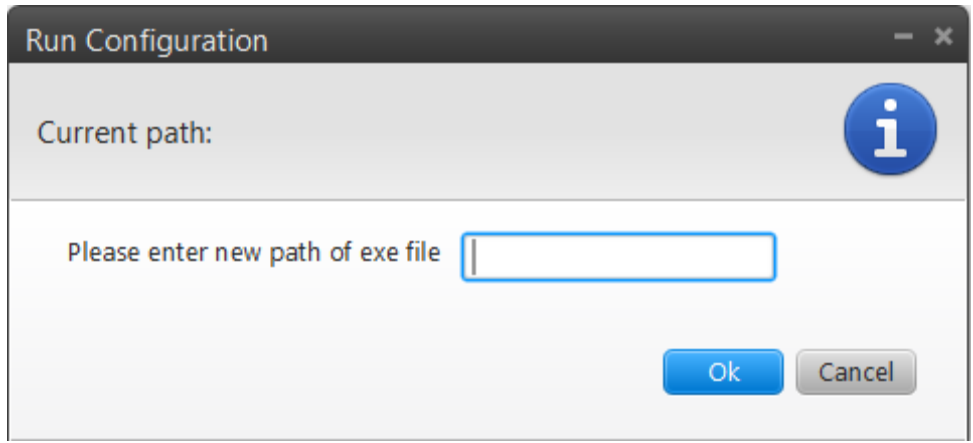

Figure 4.9     Dialog to set the path of external compiler

## 4.3   Team Collaboration

After completing above design, the designer would like to share it with his direct supervisor for approval. Therefore, Goal Net sharing function in Goal Net Designer becomes helpful. The page where user's access is managed is shown in Figure 4.10. Here, user "lisiyao" is the administrator of current Goal Net, and another user "yuhan" is being added with read access. After adding this read access, user "yuhan" will be able to open this Goal Net. However, when he opens the Goal Net, he will be notified that he only has read access to this Goal Net, and any changes will not be saved.

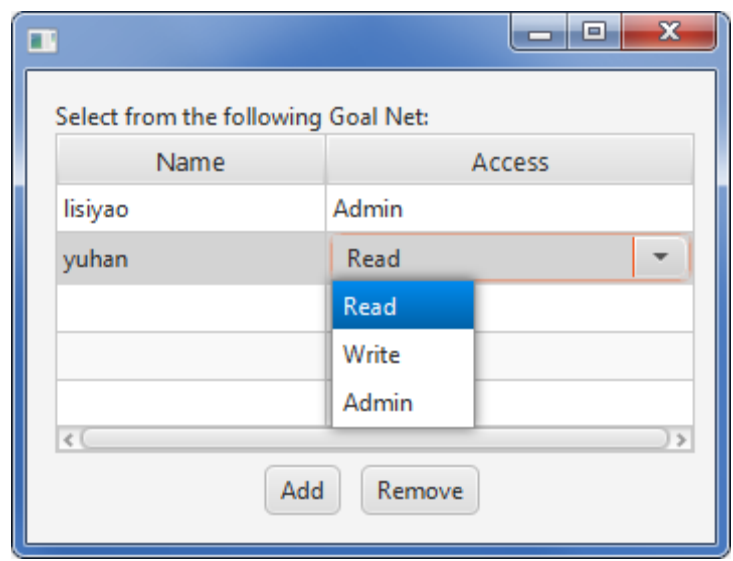

Figure 4.10     Adding user with read access to Goal Net

The designer has another task which is to model agile SDLC in another Goal Net. As there are some commonalities between waterfall SDLC and agile SDLC, some parts of existing design on waterfall SDLC can be reused. After investigation, task "Do Design" is to be cloned to and reused in new Goal Net. Figure 4.11 shows the dialog where the task and targeted Goal Net are selected.  After confirmation, the task "Do Design" and its associated functions are cloned to agile SDLC. By doing so, there is less effort required in designing the new agile SDLC Goal Net.

Finally, after direct supervisor of the designer approves Goal Net SDLC, the designer is ready to present this agent design to the entire team. To assist presentation, the design should be incorporated as part of presentation slides. In this case, the Goal Net is exported as a picture. Figure 4.12 shows the simple dialog that enables export of current Goal Net as a PNG file.

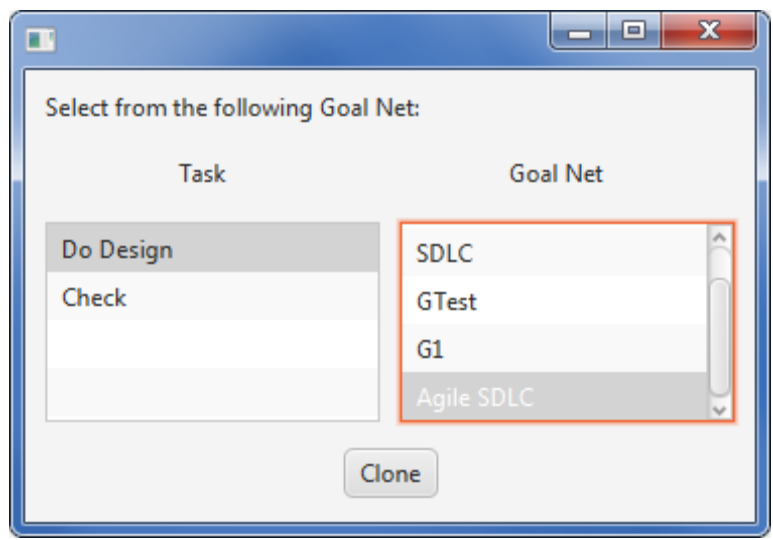

Figure 4.11     Clone task "Do Design" to Goal Net "Agile SDLC"

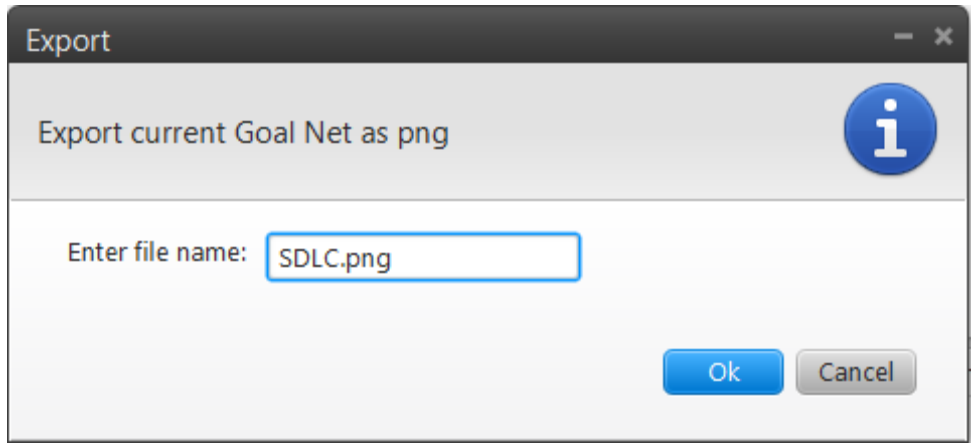

Figure 4.12    Dialog to export current design as PNG file

In summary, the process starting from the beginning of section 4 is a typical flow of agent design process using Goal Net Designer. After design phase, the job is passed to agent creator and knowledge loader to implement this agent.

# 5  CASE STUDY: TEACHABLE AGENT IN VIRTUAL REALITY

This section presents a case study on a teachable agent designed and implemented using the new Goal Net Designer. Section 5.1 introduces the background knowledge of this case study. Section 5.2 demonstrates Goal Net Designer's role in development of teachable agent. Section 5.3 describes agent implementation based on Goal Net design and the final result.

## 5.1  Background

The concept of teachable agent (TA) is motivated by Learning by Teaching theory, which stated that students learn much better when they teach their peers, compared to learning individually [19]. TA is an agent that can be taught by students, by which students' learning can be improved [20]. Interaction between TA and students is crucial for the effectiveness of entire process. Therefore, TA should avoid passively reacting to student's actions, and instead, TA should possess autonomy in learning knowledge. Such requirement enables TA to be modelled in a goal-oriented way, in which the ultimate goal of TA is to master the subject taught by student. Goal Net is adopted in this TA because it is natural to decompose the TA's ultimate goal into different sub-goals, and by achieving sub-goals, the ultimate goal can be achieved.

Furthermore, to achieve better interaction between TA and student, many TAs have been implemented in virtual reality [21] [22]. TAs in virtual reality may appear as non-player characters (NPC), and they may co-exist with some other types of agents. Essentially, such virtual reality can be viewed as a multi-agent system where agents interact among themselves and with students.

In this case study, a TAs, Little Water Molecule, is presented in a virtual world, Virtual Singapura. In this study, by interacting with this TA, students are expected to learn how water is transported among different parts of a plant. The design of this TA can be reused in various contexts, by using the same Goal Net design with different implementations and knowledge domains.

## 5.2 TA Design Using Goal Net Designer

Following the steps in Goal Net methodology, agent development starts from requirement analysis. In order to achieve better learning outcome, the TA should be able to behave in the following two ways similar to active learners in real life [23]:

- React to external events and select a goal to pursue in this circumstance
- Without external event, internally select a goal to pursue (intrinsically motivated)

With above-mentioned behaviours of a TA, it is natural to come up with the highest level Goal Net design in Goal Net Designer as shown in Figure 5.1.

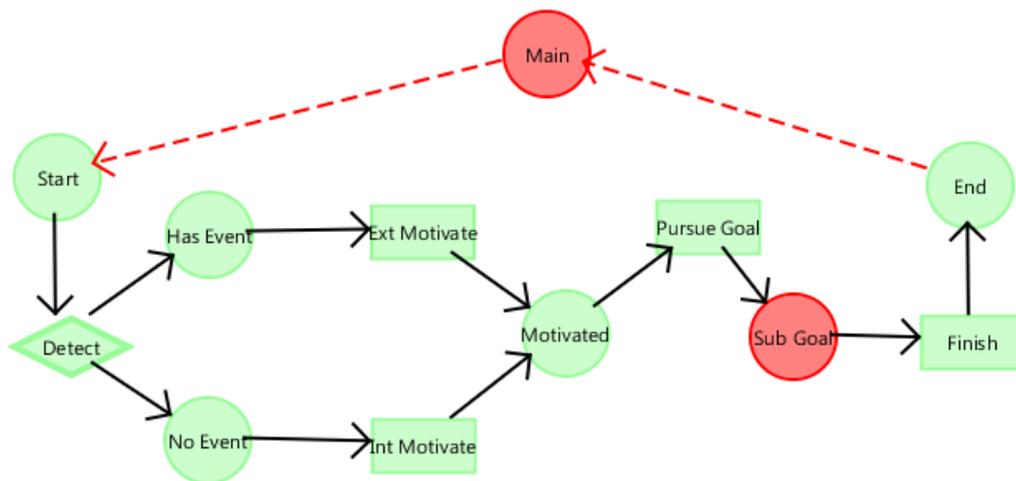

Figure 5.1    TA's main routine, exported from Goal Net Designer

The second step is to split the highest level Goal Net into sub Goal Nets. According to learning by teaching theory, TA should have the following three abilities to facilitate learning [19]:

- Teachability: TA must be able to learn new knowledge
- Practicability: TA can apply learnt knowledge in related context
- Affectivability: TA can elicit and express emotions like human being

The three abilities can be transformed into three sub-goals, namely to learn, to practice, and to be affective. Each of the three sub-goals can be modelled in another Goal Net. For example, the sub Goal Net for "to be affective" is shown in Figure 5.2, in which desire, relationship, and relevance are assessed and emotion is generated based on assessment results.

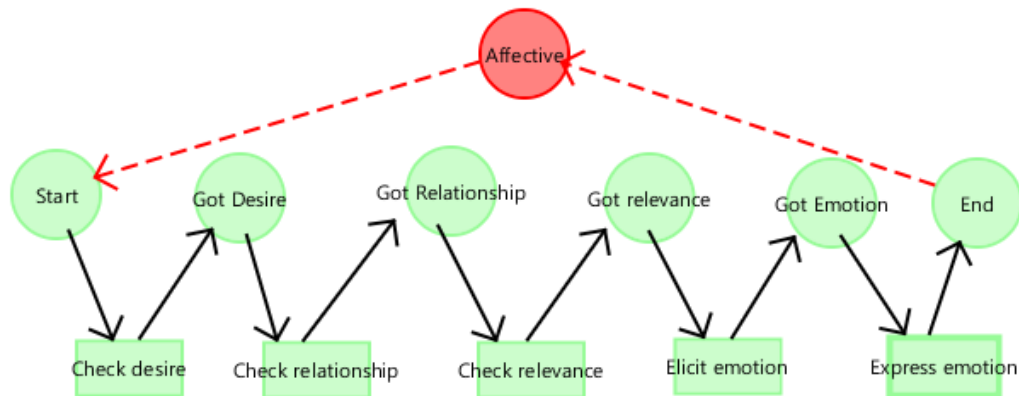

Figure 5.2    Sub Goal Net for affectivability, exported from Goal Net Designer

Since the three sub goals are part of TA's goal hierarchy, they are integrated into the main routine of TA. No splitting of TA into multiple agents is done.

The third step is detailed Goal Net design. In this step, functions are associated with states and tasks, and tasks are associated with transitions. Taking Goal Net of "to be affective" in Figure 5.2 as example, each transition is associated with a task, which contains a rule based function. Each rule based function checks the environment and past events, and returns an outcome. By combining the power of all rule-based functions, the Goal Net can generate emotion based on a series of rules.

New features in Goal Net Designer greatly enhance efficiency and productivity in the first three steps. Since these Goal Net designs require expertise in psychology, software engineering, game development, and Goal Net model, all Goal Nets are built collaboratively using team collaboration functions. It greatly reduces communication effort. Model validation tools have also identified several missing task-function associations, which can only be discovered by manual checking in the past.

Up to now, the TA has been designed in Goal Net Designer. To bring this TA to live, implementation of rule-based functions and environments, as well as GUI are still required in object-oriented programming. However, this part of implementation contains no concept of agent technology and can be done on any platform. The final step, agent implementation and its details, are discussed in the next section.

## 5.3 Agent Implementation

Based on Goal Net designed in the previous section, the TA is ready to be implemented and incorporated as part of Virtual Singapura. Programming of avatar of NPC, environment, and GUI is done using Unity. Also, as the last step of agent development, functions specified in Goal Net design of this TA is also programmed, and function interfaces are exposed so that it can be triggered externally.

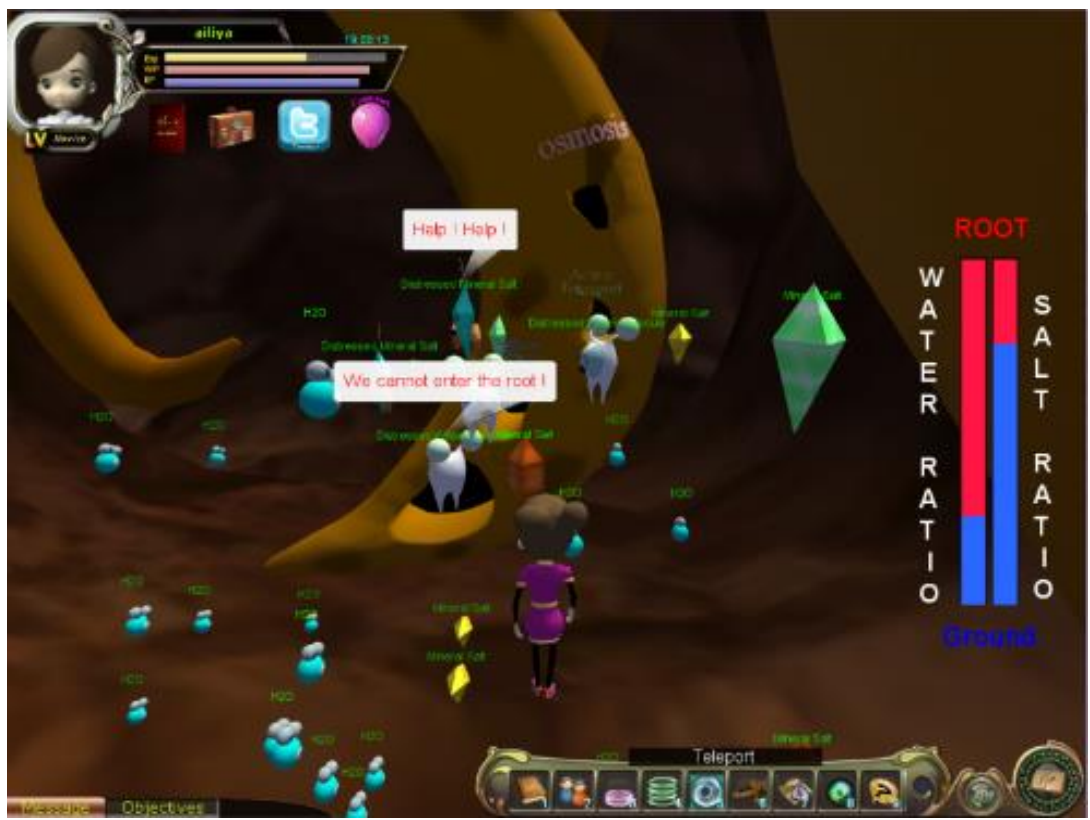

Figure 5.3   Avatar of TA, Little Water Molecule, in Virtual Singapura

After above implementation is completed, an agent is ready to be created by MADE. By triggering MADE from within Goal Net Designer, the ID of main Goal Net for TA is passed to MADE, and MADE runtime loads the Goal Net from database for agent creation. The final avatar of TA in Virtual Singapura is shown in Figure 5.3. The TA behaves in a goal-oriented way in Virtual Singapura. Figure 5.4 shows the scenario where the TA is pursuing the goal "to learn" by asking student to teach.

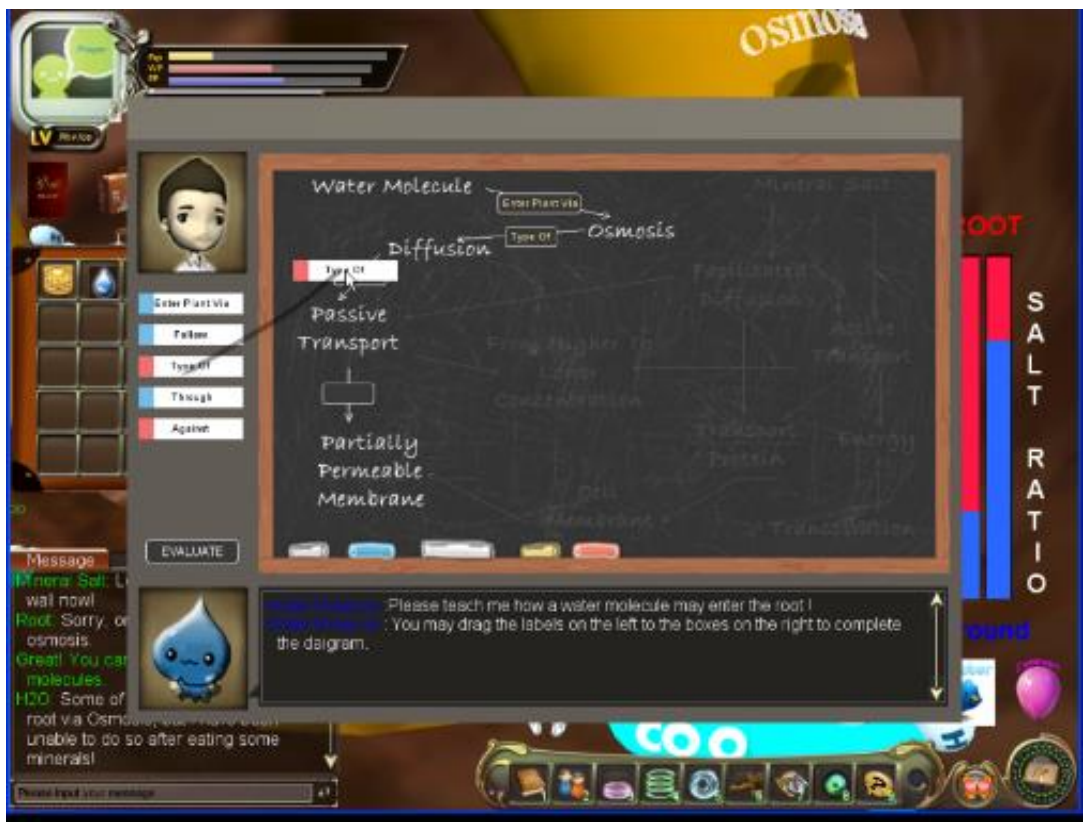

Figure 5.4   TA currently pursuing the sub-goal "to learn"

To sum up, this case study presents the whole development cycle of a TA in virtual reality using Goal Net Designer. Goal Net Designer is the starting point and plays a crucial role throughout the process. By using the new Goal Net Designer with other components in MADE, design of agent is much simplified.

# 6 CONCLUSION AND FUTURE WORK

This section concludes this report in sub-section 6.1 and provides some insights for future work in sub-section 6.2.

## 6.1 Conclusion

Agent has increasingly been recognised as a promising approach in software engineering. The proactivity and goal-orientation of agents enable problem solving in a complex environment. With research efforts on agent theories and agent development methodologies, several methodologies have been proposed and accompanying agent development tool has been developed. However, agent technology is still far from widespread application in industry due to multiple difficulties, such as prerequisite on agent-related knowledge and programming, and gaps in different phases of development.

To resolve the difficulties, a designer tool, Goal Net Designer, is built for Goal Net methodology in this project. It becomes an easy-to-use tool to model a goal-oriented agent using Goal Net model, which resembles human's goal-oriented behaviours. Goal Net Designer also successfully bridges agent design with agent implementation.

The new Goal Net Designer has been designed with better software engineering practices, included more user-friendly features, and incorporated powerful tools such as team collaboration, model validation, and link to compiler. It has also successfully accomplished the target of cross-platform and web access. Lastly, its built-in behaviour tracker and feedback system allow it to continuously evolve in future and to capture emerging needs of agent development.

The new Goal Net Designer has received good feedback from users in alpha release. Currently, Goal Net Designer is still being tested by selected users. The first release to the public can be expected within the next two months.

## 6.2 Future Work

Current work on the agent designer can still be extended in many ways, and the following paragraphs lay out some suggestions and aspects for future work.

In terms of Goal Net Designer, more enhancements can be made on functionality and features in future versions:

- Model validation tools can be enhanced to check some deeply rooted errors. Currently, Goal Net Designer is only able to check syntax errors. It is not capable of checking semantic errors like invalid function calls, infinite loops, and deadlock in Goal Net. Although this is a rather big and complicated topic, it will be very beneficial if such checks can be performed and warnings can be issued to users.
- More flexibilities and more user-friendly features can be added to Goal Net Designer to further increase the efficiency of agent development. For example, clone functions can be extended to higher level abstractions like state and transition, and more features, such as copy, cut, and paste can be explored to greatly reduce the number of operations required by user. Data collected by user behaviour logging may reveal some other aspects of improvement as well.
- Automatic link to external functions can be added. Although Goal Net Designer should not directly import functions from external sources like DLL, an automatic link to those sources from Goal Net Designer should be performed once such DLL exists. This is to reduce the efforts required in design phase, especially for designers who are not familiar with concepts such as function, return type, and parameter.
- Goal Net Designer can be closely integrated with other components in agent designer. On top of the newly implemented Run function, Goal Net Designer should provide an approach to link to external debuggers, allow monitoring of current system status, insert breakpoints, and display variables. It requires implementation that is beyond the scope of Goal Net Designer itself, too.

As the focus of this project is to enhance Goal Net Designer, there is little work done on the other two components of the agent designer, namely agent creator and knowledge loader. However, as an integrated agent design and implementation tool, they should function seamlessly with Goal Net Designer. Therefore, the following can be done to enhance the entire agent designer:

- Adapt agent creator and knowledge loader so that it is aligned with Goal Net Designer. Tremendous efforts have been made to maximise compatibility of new Goal Net Designer with other components so that MADE can still function without updating. However, the agent creator and knowledge loader will miss out benefits brought by new Goal Net Designer in the absence of a patch.
- It may be useful to explore the possibility of cross-platform features of other components of agent designer. However, complete re-implementation similar to this project should be avoided to the best effort. In the case where existing code base is easy to maintain, a third-party tool can be used to achieve cross-platform access. For example, Mono can be used to port C# program to Unix-based operating systems.
- Agent creator and knowledge loader can be adapted accordingly to include debugging capabilities that can closely function with the new Goal Net Designer, as mentioned above.

The new Goal Net Designer is subject to tests and challenges from real-life agent development projects. Although the new implementation of Goal Net Designer maintains all functionality of the previous version, and a case study has demonstrated the use of the new Goal Net Designer in real-life agent development, it may still fall short of some unforeseen user requirements. With more applications built, the new Goal Net Designer, together with other components of agent designer, will demonstrate greater value in agent development.